\documentclass[aps,prd,preprint,floats,superscriptaddress]{revtex4}

\usepackage[hypertex]{hyperref}
\usepackage{graphicx}

\newcommand{\be}{\begin{equation}}
\newcommand{\ee}{\end{equation}} 
\newcommand{\bea}{\begin{eqnarray}} 
\newcommand{\eea}{\end{eqnarray}} 
 
\begin{document} 
 
\pagestyle{plain} 
 
\title{\begin{flushright} \texttt{UMD-PP-09-041} \end{flushright}
 Natural Suppression of Proton Decay in Supersymmetric Type III Seesaw 
Models}
\author{Rabindra N. Mohapatra}
\affiliation{Maryland Center For Fundamental Physics and Department of 
Physics, University of Maryland, College Park, MD, 20742 }
 
 
\begin{abstract}
 Supersymmetric standard model (MSSM) has two sources of rapid proton 
decay: (i) R-parity breaking terms and (ii) higher dimensional Planck 
induced B-violating terms; its extensions to include neutrino masses 
via the type I seesaw mechanism need not have the first of these 
problems due to the existence of B-L as a gauge symmetry but for sure 
always have the second one. If instead, neutrino masses are 
explained in a type III seesaw extension of standard model, an
anomaly free gauge symmetry different from B-L is known to exist. In 
this note, it is shown that a realistic supersymmetric versions of this 
model can be constructed (MSSM as well as SUSY left-right with type III 
seesaw) which eliminate R-parity violating couplings and  
suppress Planck scale contributions to proton decay. The degree of 
suppression of the latter depends on the weak gauge group. For the 
left-right case, the suppression to the desired level is easily achieved.
 \end{abstract}
 
\pacs{} \maketitle

 
\section{Introduction} 
Supersymmetry provides a very attractive way to solve the gauge 
hierarchy problem of the standard model. This and the fact that it also 
provides a 
natural candidate for the dark matter of the universe has made 
supersymmetric extensions of the standard model (MSSM) a prime focus of 
theoretical and experimantal research in the past two decades. Despite 
these attractive features, MSSM as a complete theory is not satisfactory 
since it takes a step backward from standard 
model as far as our understanding of proton decay is concerned. There are 
two new sources of rapid proton decay in MSSM, which arise from the fact 
that the theory has supersymmetry: (i) renormalizable R-parity breaking 
interactions, a 
combination of two of which, $QLd^c$ and $u^cd^cd^c$, lead to extremely 
rapid p-decay and secondly (ii) Planck scale induced higher dimensional 
operators 
of the form $QQQL/M_{Pl}$ and $u^cu^cd^ce^c/M_{Pl}$, which also do the 
same. Current lower limits on the 
proton lifetime can be used to set limits on the product of the two 
couplings in case (i) $\lambda'\lambda''< 10^{-24}$ and the individual 
couplings in case (ii) $\lambda < 10^{-7}$\cite{pdecay}. These are 
extremely 
stringent bounds and before supersymmetric models can be 
accepted as good descriptions of nature, one must find a satisfactory way 
to either eliminate or suppress these interactions in a natural manner. 

It was pointed out many years ago that extending MSSM to incorporate 
neutrino masses via type I seesaw mechanism (i.e. include three right 
handed neutrinos into MSSM) automatically extends the local symmetry to 
 $SU(2)_L\times U(1)_{I_{3R}}\times U(1)_{B-L}$ which when broken by 
$B-L=2$ Higgs fields automatically eliminates the undesirable R-parity 
breaking terms\cite{RNM}. However they do not eliminate the Planck scale 
induced terms and usually, one invokes additional gauged discrete 
symmetries to achieve that goal\cite{gauged}. While B-L symmetry is  
physically quite 
well motivated, symmetries\cite{gauged} used to eliminate the $QQQL$ type 
operators are often quite adhoc. We explore this question in the context 
of type III seesaw models\cite{type3}.

 In several papers\cite{barr,ma}, it has been pointed out that 
nonsupersymmetric type III  seesaw models admit an extra $U(1)_X$ 
gauge symmetry. The presence of this additional gauge symmetry provides 
a natural way to understand why the seesaw scale (in this case the 
Majorana mass of the triplet fermions of the type III seesaw model), is 
much less than the its natural value, the Planck scale in much the same 
way that the B-L symmetry provides an explanation does the same for the 
type I 
seesaw case. However as already noted, in the type I seesaw case, B-L 
symmetry while 
forbidding the R-parity breaking terms still allows the Planck scale 
operators of type $QQQL$ and $u^cu^cd^ce^c$. In this brief note 
we point out that the type III seesaw case is very different: the same 
symmetry that explains the smallness seesaw scale compared to the Planck 
scale also automatically forbids both the R-parity breaking as well as 
the Planck induced proton decay operators in the symmetry limit. When the 
gauge 
symmetry is broken, the resulting higher dimensional B-violating 
operators are suppressed with the degree of suppression depending on the 
nature of the electroweak gauge group: for instance we find that the 
required suppression is more easily achieved for 
left-right gauge group than the SM case. 

This paper is organized as follows: in sec. 2, we discuss the extension 
of MSSM with type III seesaw (called MSSM$_{III}$); in sec. 3, we discuss 
the left-right extension and in sec. 4, we conclude with some comments 
and a summary.

\section{ MSSM$_{III}$: MSSM extended with type III seesaw}
The basic idea 
of the type III seesaw model is to add three hypercharge neutral triplet 
fermions to the standard model\cite{type3}. We denote the triplets as:
\begin{eqnarray} 
\Sigma~=~\left(\begin{array}{c}t^+ \\ t^0 
\\t^-\end{array}\right)
\end{eqnarray}
 In order to find the extra 
gauge symmetry of the model, one looks at the constraints on $U(1)_X$ 
charges of the various fields from the requirement of vanishing of the 
following anomalies involving the $U(1)_X$ symmetry: (i) $U(1)_X[SU(2)_L]^2$,
(ii) $U(1)_X[SU(3)_c]^2$ (iii) $U(1)_X[U(1)_Y]^2$ (iv) 
$U(1)_Y[U(1)_X]^2$ and $U(1)_X[Gravity]^2$. These constraints for the 
nonsupersymmetric type III model has already been studied in 
\cite{barr,ma} and further analyzed in \cite{franco} where it has been 
shown that only for three triplets, there is a solution to the anomaly 
equations. The resulting charges for the various SM fields are shown in 
the first six rows of table I. They are expressed in terms of two 
arbitrary numbers $n_q$ and $n_l$, which denote the quark and lepton 
doublet charges under $U(1)_X$. We assume throughout this paper that 
$U(1)_X$ is generation blind.
 
In order to construct 
a realistic supersymmetric extension of this model, we need two 
Higgs doublet superfields,
$H_{u,d}$ and add two singlet superfields $S,\bar{S}$. Since their 
contributions to the anomaly equations cancel, the anomaly constraints 
are satisfied for the MSSM$_{III}$ case also for the same charges of 
matter fields as in \cite{barr, ma}. The $U(1)_X$ 
charges of all the fields are given in Table I.

\vspace{0.3in}
\begin{center}
\begin{tabular}{|c||c|}\hline Fields & $U(1)_X$ quantum number 
\\\hline       Q & $n_q$ \\ $u^c$ & $-\frac{1}{4}(7n_q-3n_l)$\\
$d^c$ &  $-\frac{1}{4}(n_q+3n_l)$ \\ $L$ & $n_l$ \\ $e^c$ &  
$-\frac{1}{4}(-9n_q +5n_l)$ \\ $\Sigma$ &  $-\frac{1}{4}(3n_q+n_l)$\\ 
$H_u$ &  $\frac{3}{4}(n_q-n_l)$ \\ $H_d$ &  $-\frac{3}{4}(n_q-n_l)$\\
$S$ & $+\frac{1}{2}(3n_q+n_l)$ \\ $\overline{S}$ 
&$-\frac{1}{2}(3n_q+n_l)$\\
\hline\end{tabular}
\end{center}
\vspace{0.3in}

\noindent Table Caption: Anomaly free quantum numbers for various fields 
in MSSM$_{III}$.
\vspace{0.3in} 

Note that this symmetry allows the Yukawa coupling part of 
the superpotential of the form:
\begin{eqnarray}
W_{III}~=~h_uQH_uu^c~+h_d 
QH_dd^c~+~h_lLH_de^c\frac{\overline{S}}{M_{Pl}} 
  h_\nu LH_u\Sigma ~+~f\Sigma\Sigma S
\end{eqnarray}
For the Higgs part of the superpotential, we have
\begin{eqnarray}
W_H~=~\mu H_uH_d + Y(S\overline{S}-M^2)
\end{eqnarray}
where $Y$ is a SM singlet and $U(1)_X$ neutral fields.
A curious feature of this superpotential is that the charged lepton 
masses which arise 
from the $h_l$ term in the superpotential are suppressed by the 
factr $\frac{<\bar{S}>}{M_{Pl}}$. Note however that $<S>$ and 
$<\bar{S}>$  also give mass to the $\Sigma$ 
field and  implement the type 
III seesaw mechanism. These considerations restrict the value of $<S>$. 

To reproduce the charged lepton masses (mainly the tau lepton) , it seems 
that we must choose, $<S>=<\overline{S}>=M_U\sim 3\times 10^{15}-
10^{16}$ GeV (using $M_{Pl}\sim 1.22\times 10^{18}$ GeV) for a tau Yukawa 
coupling from 3-1. This says that 
the charged lepton masses are 
automatically suppressed compared to quark masses as well as the Dirac 
mass of
the neutrinos. Also the observed neutrino masses of order $0.05$ eV or 
less imply that the triplet fermion masses doing seesaw should of order 
$10^{14}-10^{15}$ GeV or so for $h_\nu\sim 1-3$. This will mean that 
$f\sim 0.1-0.01$. The model can reproduce all observed fermion masses. As 
we show 
below, the above $<S>$ limits the degree of suppression of the p-decay 
operators in MSSM$_{III}$. It could of course be that the lepton Yukawa 
coupling term is scaled by some new physics scale below the Planck scale. 
In that case the vev$<S>$ could be lot lower, again with implications for 
the degree of suppression of the proton decay operators.

 Note incidentally that if indeed the lepton Yukawa term is scaled by 
$M_{Pl}$, the subsequent large 
value of the triplet mass will only slightly disturb the coupling 
unification of MSSM somewhat. On the other hand, if the scale is set by 
lower mass scale, the effect on gauge coupling running will affect 
unification. However as we remark below, these models do not grand unify 
in the usual manner to $SU(5)$ or $SO(10)$ groups and still keep the extra 
$U(1)_X$ gauge 
symmetry. 

\subsection{Suppression of baryon violating operators}
Using the $U(1)_X$ charge assignments for fields in the Table I, we 
compute the $U(1)_X$ charges of the 
baryon and lepton number violating operators and list them in Table 2 
below. 
\vspace{0.3in} 
\begin{center}
\begin{tabular}{|c||c|}\hline\hline
$QQQL$ & $3n_q+n_l$\\ $u^cu^cd^ce^c$ & $-\frac{1}{2}(3n_q+n_l)$\\
$QLd^c$ & $\frac{1}{4}(3n_q+n_l)$ \\ $LLe^c$ & $\frac{3}{4}(3n_q+n_l)$\\
$u^cd^cd^c$ & $-\frac{3}{4}(3n_q+n_l)$\\ $LH_uLH_u$ & 
$\frac{2}{4}(3n_q+n_l)$\\ $LH_u$ & $\frac{1}{4}(3n_q+n_l)$\\
\hline \end{tabular}
\end{center}
\vspace{0.3in}

\noindent Table Caption: $U(1)_X$ charges of the various baryon and 
lepton number violating operators.
\vspace{0.3in} 

Note that all the R-parity breaking and higher dimensional proton decay 
operators have 
$U(1)_X$ charges which are multiples of the $U(1)_X$ charge of the 
triplet field $n_\Sigma =
-\frac{1}{4}(3n_q+n_l)$ . We require the latter to be nonzero so that 
the smallness of the triplet mass 
compared to the Planck mass can be understood. As a result, we find that 
all R-parity breaking as well as $QQQL$ operators are forbidden to all 
orders.  As far as the $u^cu^cd^ce^c$ operator goes, while it is forbidden 
in the symmetry limit, it is induced when $U(1)_X$ symmetry is 
broken by the operator $u^cu^cd^ce^cS/M_{Pl}$; if we choose 
$<S>=3\times 10^{15}$, the suppression is about $3\times 10^{-3}$ and one 
still needs to 
dial the coupling of this operator down by a factor of a $3\times 10^{-5}$ 
or so 
to keep it compatible with current proton lifetime limits.  The 
$QQQL$ operator is induced at the next higher order i.e. 
$QQQL\bar{S}^2/M^2_{Pl}$ so that after symmetry breaking, its strength 
is $\sim 10^{-5}$ and one needs a coupling for this operator of about 
$10^{-2}$. Thus there is suppression for these operators but not 
enough due to lepton mass constraint. It does however ameliorate the fine 
tuning problem somewhat. As far as the 
R-parity violating operators are concerned though, they are forbidden to 
all orders.

Note that if the lepton Yukawa is scaled by a lower mass scale, the $<S>$ 
induced  proton decay operators are more easily suppressed. 

\section{``Type III seesaw" with left-right symmetry}
In this section, we extend our analysis to the case where the electroweak 
gauge group is the left-right symmetric group $G_{LR}\equiv SU(2)_L\times 
SU(2)_R\times U(1)_{B-L}$. Left right models with ``type III seesaw"
 have been considered in\cite{perez}. We will study 
the susy version of this model and discuss the p-decay operators.
The full gauge group will be $G_{LR}\times U(1)_X$, where $U(1)_X$ is the 
new anomaly 
free gauge group arising from type III seesaw as we show below.
In this case, we choose the quark and lepton assignment to the LR 
group as follows:
\begin{eqnarray}
Q~=~\left(\begin{array}{c}u \\ d\end{array}\right)(2,1,1/3);~~  
Q^c=\left( 
\begin{array}{c}d^c\\u^c\end{array}\right)(1,2,-1/3);\\ \nonumber
L~=~\left(\begin{array}{c}\nu 
\\ 
e\end{array}\right)(2,1,-1); ~~
 L^c~=~\left(\begin{array}{c}e^c\\\nu^{c}\end{array}\right)(1,2,+1)
\end{eqnarray}
The $U(1)_X$ charges are given by the numbers in the parenthesis next to 
the field in what follows: $Q(n_q)$;  $Q^c(-n_{q^c})$;  $L(n_l)$;
 ${L^c}(-n_{l^c})$.
The left-right symmetric assignment of the triplets required for the type 
III seesaw are given by:
\begin{eqnarray}
\Sigma^c~=~\left(\begin{array}{c}T^+\\ 
T^0\\T^-\end{array}\right)(1,3,0,-n_{\Sigma^c});\\ 
\nonumber
\Sigma~=~\left(\begin{array}{c}t^+\\t^0\\t^-\end{array}\right)
(3,1,0,n_{\Sigma})
\end{eqnarray}
where the $U(1)_X$ charges are included above.
There are six anomaly 
conditions in this case and a solution to the vanishing of all the 
anomaly constraints is:
\begin{eqnarray}
n_{q}=n_{q^c}\equiv n_q;~~n_{l}=n_{l^c}\equiv n_l: 
~~n_{\Sigma}=n_{\Sigma^c}=-\frac{1}{4}(3n_q+n_l)
\end{eqnarray}
In order to implement symmetry breaking, this model will have left and 
right Higgs 
doublets which will come in pairs so that their anomalies will cancel 
among themselves: we denote them by $\chi (2,1,+1), \bar{\chi}(2,1,-1); 
\chi^c (1,2,-1), 
\bar{\chi}^c (1,2,+1)$. The $U(1)_X$ quantum numbers for $\chi$ and ,
$\bar{\chi}$ are $\pm\frac{3}{4}(n_q-n_l)$ respectively and 
 opposite for the $\chi^c$ and $\bar{\chi}^c$ fields. 
We also have a bi-doublet $\phi (2,2,0)$ with zero 
$U(1)_X$ quantum number in the model. As in the MSSM$_{III}$ case, here 
have two singlet fields $S$ ans 
$\bar{S}$ that have $U(1)_X$ quantum number of $\pm\frac{1}{2}(3n_q+n_l)$
The Yukawa coupling superpotential for this model is given by:
\begin{eqnarray}
W^{LR}_{Y}~=~h_qQ\phi Q^c+h_lL\phi L^c+ f 
(L\chi\Sigma+L^c\chi^c\Sigma^c)
+\lambda(\Sigma^2S+\Sigma^{c2}\bar{S})
\end{eqnarray}
The different symmetry breaking stages are as follows: We have  
 $<S>~=~<\bar{S}>~\sim~ <\chi^c> \gg <\phi>$. The first breaks the 
$U(1)_X$ 
symmetry and give large masses to the triplet fields. $<\chi^c>$ breaks 
the left right group down to the standard model symmetry and gives mass 
term of the form $\nu^c T^0$ and  $e^cT^-$ field.
  The bi-doublet vev which breaks the standard model group and gives mass 
to the charged  fermions as well as the quarks. The vev of the $\chi$ field 
is assumed to be zero. 

The neutrino mass in this model arises 
from a double seesaw mechanism involving the fields $(\nu, \nu^c, 
T^0)$ as discussed in \cite{perez}. 
Strictly speaking, since in this model, the left triplet fermion does not 
participate in the seesaw mechanism, it is not a canonical type III 
seesaw model; it is more appropriate to call it the triplet version of 
double seesaw mechanism\cite{valle}. 

A fundamental difference between this model and the MSSM$_{III}$ 
discussed in section 2 is that the $<S>$ vev in this model need not be at 
the GUT scale and in fact, since neutrino mass involves the double 
seesaw, this vev can be anywhere between tens of TeV range to $10^{14}$ 
GeV. This has implications for the suppression of R-parity and baryon 
violating operators. 

\subsection{R-parity and higher dimensional proton decay operators}
Let us now discuss the R-parity and baryon number violating operators in 
this model. In the usual susy left-right case with doublets breaking 
B-L symmetry, there 
are tree level R-P violating operators of the form $L^c\chi^c$ and 
nonrenormalizable ones of the form 
$LLL^c\bar{\chi}^c$ and $Q^cQ^cQ^c\bar{\chi}^c$. However 
in this model, these operators are forbidden by $U(1)_X$ gauge symmetry 
since their $U(1)_X$ charges respectively are: $-\frac{1}{4}(3n_q+n_l)$,
$+\frac{1}{4}(3n_q+n_l)$, $-\frac{3}{4}(3n_q+n_l)$. Again as in the 
MSSM$_{III}$ case, these charges are 
all proportional to the triplet charges and therefore as long as the 
triplet charges are nonvanishing, which we require to understand their 
masses being smaller than the Planck mass, the R-parity breaking  
operators are forbidden.

Turning to higher dimensional proton decay operator, we note that the 
$QQQL$ and $Q^cQ^cQ^cL^c$ operators are forbidden by $U(1)_X$ symmetry.
The leading order $U(1)_X$ invariant operator 
that contributes to $QQQL$ and $Q^cQ^cQ^cL^c$ after symmetry breaking are 
$QQQL\bar{S}^2/M^2_{Pl}$  giving strength of 
these operators to be $(<\bar{S}>/M_{Pl})^2$. For 
$<\bar{S}>$ vev below $10^{12}$ GeV, this is fully consistent with 
current limits discussed in the introduction. Thus the model is safe with 
respect to rapid proton decay due to the presence of $U(1)_X$ symmetry.

\section{Comments and Conclusion}
  In summary, in this brief note, we have analysed and pointed out a novel 
feature of the extra gauged  $U(1)_X$ symmetry which is present in 
type III seesaw extensions of standard model for explaining small 
neutrino masses. First we show that this symmetry remains in the 
supersymmetric extensions of the simple type III model, MSSM$_{III}$ as 
well as its left-right symmetric extension. These models provide 
realistic descriptions of nature and  more importantly, 
 they not only forbid the R-parity breaking interactions of 
MSSM but they also suppress 
the undesirable and dangerous Planck scale induced proton decay 
interactions. The suppression of the latter 
operators to the desired level is more easily achieved in the 
supersymmetric 
left-right versions of the model than in the MSSM$_{III}$. These 
results eliminate a 
conceptual problem of MSSM and its generalizations to include neutrino 
masses and puts 
type III seesaw models at an advantage over the type I seesaw 
models. However we also 
find that these models fail to grand unify to conventional SU(5) or 
SO(10) groups while keeping the extra $U(1)_X$, unlike the type I seesaw 
models which naturally grand unify to SO(10) (although an extension to the 
case of $SU(2)_L\times SU(2)_R\times SU(4)_c \times U(1)_X$ seems straight 
forward).

 This work is supported by the US National Science Foundation under grant 
No. PHY-0652363 and Alexander von Humboldt Award (2005 Senior Humboldt 
Award). The 
author is grateful to Manfred Lindner for hospitality at the Max Planck 
Institut fur Kernphysik in Heidelberg during the time when the work was 
completed.

\end{document}